\documentclass[aps,prl,twocolumn,superscriptaddress,numerical]{revtex4-2}
\usepackage[utf8]{inputenc}
\usepackage[english]{babel}
\usepackage[T1]{fontenc}
\usepackage{amsmath,amssymb,dsfont}
\usepackage[colorlinks,
citecolor=blue,
linkcolor=blue,
urlcolor=magenta]{hyperref}
\usepackage{cases, bm}

\usepackage{orcidlink}
\usepackage{multirow}
\usepackage{mathrsfs}
\usepackage[normalem]{ulem}

\newcommand{\prlsection}[1]{\textbf{\textit{#1.}}}

\begin{document}

\title{Steady-state density large deviations in driven diffusive systems}

\author{Soumyabrata Saha~\orcidlink{0009-0006-7539-8801}}
\affiliation{Department of Theoretical Physics, Tata Institute of Fundamental Research, Homi Bhabha Road, Mumbai 400005, India.}

\author{Tridib Sadhu~\orcidlink{0000-0003-0390-6257}}
\affiliation{Department of Theoretical Physics, Tata Institute of Fundamental Research, Homi Bhabha Road, Mumbai 400005, India.}

\begin{abstract}
A diffusive system coupled to boundary reservoirs reaches a non-equilibrium steady state whose density fluctuations can exhibit correlations over macroscopic distances. While current fluctuations are well understood for generic systems, exact results for the steady-state density large-deviation functional remain restricted to a few special models. We introduce a field transformation within Macroscopic Fluctuation Theory that, in one dimension, captures two broad exactly solvable classes characterized respectively by local and non-local large-deviation functionals. The transformation also allows us to explicitly determine the optimal fluctuation paths, exposing the dynamical origin of non-locality and time irreversibility. For generic transport coefficients, the same transformation provides a systematic perturbative scheme in the boundary drive. In particular, we show that a finite bulk drive shifts the onset of non-locality from quadratic to linear order in the boundary drive and generically generates long-range connected multipoint correlations of arbitrarily high order. In higher dimensions, this onset depends on the orientation of the bulk drive relative to the boundary-induced density gradient.
\end{abstract}

\maketitle

\prlsection{Introduction} Boundary-driven systems provide a paradigmatic setting for characterizing a \emph{non-equilibrium steady state} (NESS). A widely studied realization consists of an extended diffusive system coupled at its boundaries to unequal reservoirs~\cite{2007_Derrida_Non,2015_Mallick_The}. While the bulk dynamics locally conserves a quantity, such as particle number or energy, the reservoirs exchange this conserved quantity with the system, thereby sustaining a non-zero average current through the system. This persistent transport can generate correlations over macroscopic distances and break the time-reversal symmetry of fluctuations, making the NESS qualitatively different from its equilibrium counterpart.

Much of our understanding of such systems has come from fluctuations of the transported quantity, in particular the statistics of the time-integrated current~\cite{2004_Bodineau_Current,2005_Bertini_Current,2006_Bertini_Non,2007_Bodineau_Cumulants}. A complementary characterization is provided by fluctuations of the instantaneous spatial distribution of the conserved quantity, namely its density profile~\cite{2007_Derrida_Non,2015_Bertini_Macroscopic}. For a large one-dimensional system of size $L$ maintained in a NESS, the probability of observing a coarse-grained density profile $r(x)$ obeys the large-deviation form~\cite{2015_Bertini_Macroscopic}
\begin{equation}\label{eq:ldf_defn}
\Pr(r(x))\asymp\mathrm{e}^{-L\psi(r(x))}
\end{equation}
where $\psi(r(x))$ is the large-deviation functional (ldf).

The ldf provides a natural non-equilibrium analogue of the equilibrium free-energy functional~\cite{2015_Mallick_The,2021_Derrida_Large}. Beyond quantifying the exponential cost of a macroscopic fluctuation, it characterizes how different regions of the system fluctuate collectively and thereby encodes the spatial correlations of the NESS. Its structure can further reveal emergent symmetries as well as signal phase transitions. Determining the LDF is, however, challenging, with numerical approaches generally requiring specialized rare-event methods~\cite{2006_Giardina_Direct,2007_Lecomte_A}, while exact analytical results remain limited to a few special cases~\cite{2001_Derrida_Free,2002_Derrida_Large,2002_Derrida_Exact,2003_Derrida_Exact,2004_Enaud_Large,2005_Bertini_Large,2021_Derrida_Large}. A general understanding of when the ldf becomes non-local, as well as of the dynamical paths generating a prescribed fluctuation, therefore remains incomplete~\cite{2004_Bodineau_Current,2007_Derrida_Non}.

Macroscopic fluctuation theory (MFT) provides a more general route by describing diffusive systems at hydrodynamic scales in terms of two density-dependent transport coefficients, the diffusivity and mobility~\cite{1991_Spohn_Large,2004_Bodineau_Current}. This coarse-grained description has both reproduced results obtained from exact microscopic approaches~\cite{2001_Bertini_Fluctuations,2002_Bertini_Macroscopic,2007_Tailleur_Mapping,2008_Tailleur_Mapping,2011_Derrida_Microscopic,2022_Mallick_Exact,2024_Mallick_Exact,2024_Saha_Large} and yielded predictions beyond microscopically tractable regimes~\cite{2005_Bertini_Large,2014_Krapivsky_Large,2017_Baek_Dynamical,2023_Agranov_Tricritical,2023_Dandekar_Dynamical,2023_Agranov_Macroscopic,2024_Grabsch_Tracer,2024_Sharma_Large,2025_Berlioz_Tracer,2025_Grabsch_Exact}. Determining the density LDF within MFT nevertheless requires solving a nonlinear variational problem, which has so far been exactly tractable only in a few cases, typically through model-specific constructions.

In this \emph{Letter}, we show that these seemingly distinct solvable cases can be brought under a common formalism through a suitable transformation of the MFT fields. This recovers the known LDFs while explicitly constructing the optimal trajectories and exposing the dynamical origin of long-range correlations and time-irreversible fluctuations. For cases that do not admit an exact solution, the same transformation leads to a systematic perturbative approach that yields the leading non-local LDF and reveals that a finite bulk drive shifts the onset of non-locality from quadratic to linear order in the boundary drive. Both the transformation and the perturbative approach extend naturally to higher dimensions, where they capture how the onset of non-locality and the optimal fluctuation dynamics depend on the orientation of the bulk drive.

\prlsection{Setup} A diffusive system with one conserved quantity admits a hydrodynamic description on length-scales $\ell$ much larger than microscopic lengths and on time-scales $\ell^2$ much longer than microscopic relaxation times, with the coarse-grained density of the conserved quantity emerging as the relevant slow field. Introducing the hydrodynamic coordinates $x=\frac{X}{\ell}$ and $t=\frac{T}{\ell^2}$, this density field $\rho(x,t)$ obeys the fluctuating hydrodynamic equation~\cite{2007_Derrida_Non,2011_Derrida_Microscopic,2026_Saha_A}
\begin{equation}\label{eq:fhd_eqn}
\partial_t\rho=\partial_x\bigg(D(\rho)\partial_x\rho-\frac{\nu}{2}\sigma(\rho)+\frac{1}{\sqrt{\ell}}\sqrt{\sigma(\rho)}\eta\bigg).
\end{equation}
Here, the diffusivity $D(\rho)$ governs the diffusive relaxation of density gradients, while the mobility $\sigma(\rho)$ controls both the response to a bulk driving field $\nu$ and the amplitude of the fluctuating component of the current, with $\eta(x,t)$ denoting a Gaussian white noise of zero mean and unit variance. The factor $\frac{1}{\sqrt{\ell}}$ reflects the suppression of these fluctuations under coarse-graining. The model-dependent transport coefficients are related to each other through the fluctuation-dissipation relation~\cite{1977_Hohenberg_Theory,1991_Spohn_Large} $\frac{2D(\rho)}{\sigma(\rho)}=f''(\rho)$, where $f(\rho)$ is the canonical free-energy density. Since the local chemical potential is given by $\mu(\rho)=f'(\rho)$, this equivalently yields the relation $\mu'(\rho)=\frac{2D(\rho)}{\sigma(\rho)}$.

\begin{figure}
\centering
\includegraphics[width=0.99\linewidth]{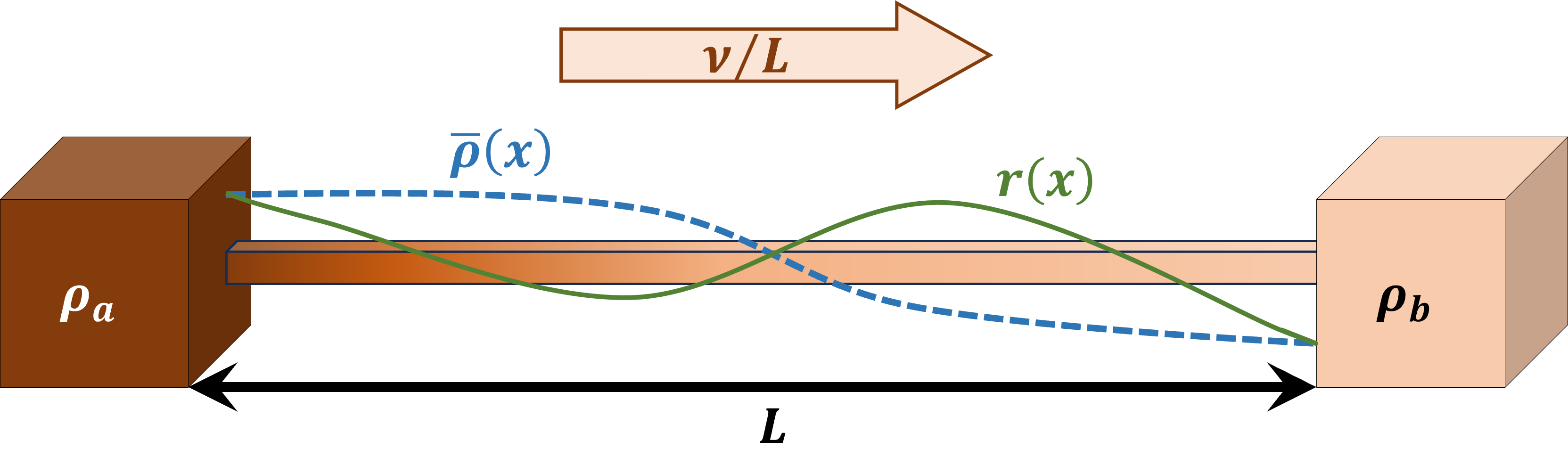}
\caption{A diffusive system of length $L$ coupled to reservoirs of densities $\rho_a$ and $\rho_b$, with a weak bulk drive $\frac{\nu}{L}$. The typical NESS profile is denoted by $\bar{\rho}(x)$, while $r(x)$ denotes a density fluctuation.}
\label{fig:system_scheme}
\end{figure}

Our focus in this \emph{Letter} is a one-dimensional system of length $L$ coupled to reservoirs that fix the boundary densities at $\rho_a$ and $\rho_b$, with their difference $\rho_a-\rho_b$ providing the boundary drive that sustains a density gradient. We additionally consider a weak bulk bias $\frac{\nu}{\ell}$, arising either from an external forcing or an intrinsic microscopic asymmetry, and scaled so that $\nu$ remains finite in the hydrodynamic current. A schematic of this setup is shown in fig.~\ref{fig:system_scheme}. Since we probe density fluctuations on the scale of the full system, we henceforth set $\ell=L$, so that $x\in[0,1]$. At long times, this system reaches a NESS characterized by a spatially uniform average current $\bar{J}$ that obeys the generalized Fick's law
\begin{equation}
\bar{J}=-D(\bar{\rho})\bar{\rho}'+\frac{\nu}{2}\sigma(\bar{\rho})={\rm constant},
\end{equation}
with $\bar{\rho}(x)$ denoting the typical density profile in the steady state. Together with the boundary conditions $\bar{\rho}(0)=\rho_a$ and $\bar{\rho}(1)=\rho_b$, the above determines $\bar{\rho}(x)$. The ldf $\psi(r(x))$ describes rare density fluctuations around this typical profile.

\prlsection{MFT framework} Within MFT, the probability of observing a density fluctuation $r(x)$ is obtained by considering all possible histories from the typical profile $\bar{\rho}(x)$ far in the past ($t\to-\infty$) to $r(x)$ at the observation time ($t=0$). The fluctuating hydrodynamics in \eqref{eq:fhd_eqn} assigns a probability weight to each such history, which can be written as a Martin–Siggia–Rose–Janssen–De-Dominicis (MSRJD) path-integral~\cite{1973_Martin_Statistical,1976_Janssen_On,1976_Dominicis_Techniques,1978_Dominicis_Field}
\begin{subequations}\label{eq:mft_defn}
\begin{equation}\label{eq:mft_path_int}
\Pr(r(x))=\int_{\bar{\rho}(x)}^{r(x)}[\mathcal{D}\widehat{\rho},\mathcal{D}\rho]\,\mathrm{e}^{-L\int_{-\infty}^{0}\mathrm{d}t\left[\int_{0}^{1}\mathrm{d}x\left(\widehat{\rho}\partial_t\rho\right)-H(\widehat{\rho},\rho)\right]}
\end{equation}
where $\widehat{\rho}$ is the response field arising due to the local conservation law. The effective Hamiltonian is given by
\begin{equation}\label{eq:generic_hamilton}
H(\widehat{\rho},\rho)=\int_{0}^{1}\mathrm{d}x\,\bigg(\frac{\sigma(\rho)}{2}\partial_x\widehat{\rho}-D(\rho)\partial_x\rho+\frac{\nu}{2}\sigma(\rho)\bigg)\partial_x\widehat{\rho}
\end{equation}
\end{subequations}
is expressed in terms of the two system-dependent transport parameters.

For large $L$, the path-integral in \eqref{eq:mft_path_int} is dominated by the path of minimal-action~\cite{2007_Derrida_Non,2015_Bertini_Macroscopic}, which is the coefficient of $L$ in the exponential. This leads to the large-deviation asymptotic for $\Pr(r(x))$, with the ldf $\psi(r(x))$ given by the minimal-action. The optimal fields that minimize the action are solutions of the Euler-Lagrange equations~\cite{2009_Derrida_Current2,2014_Krapivsky_Large,2022_Mallick_Exact},
\begin{subequations}\label{eq:old_el_eqns}
\begin{align}
\partial_t\widehat{\rho}&=-D(\rho)\partial_x^2\widehat{\rho}-\frac{\sigma'(\rho)}{2}\partial_x\widehat{\rho}\big(\nu+\partial_x\widehat{\rho}\big) \label{eq:old_el_i}\\
\partial_t\rho&=\partial_x\bigg[D(\rho)\partial_x\rho-\sigma(\rho)\bigg(\frac{\nu}{2}+\partial_x\widehat{\rho}\bigg)\bigg] \label{eq:old_el_ii}
\end{align}
\end{subequations}
subjected to the boundary conditions
\begin{subequations}\label{eq:opt_bc}
\begin{align}
&\widehat{\rho}(0,t)=\widehat{\rho}(1,t)=0,\label{eq:rhohat_spatial_bc}\\
&\rho(0,t)=\rho_a,\;\rho(1,t)=\rho_b,\label{eq:rho_spatial_bc}\\
&\rho(x,-\infty)=\bar{\rho}(x),\;\rho(x,0)=r(x).
\end{align}
\end{subequations}

The initial steady state is a fixed point of the Euler-Lagrange equations, where $\widehat{\rho}=0$ and hence $H=0$. Since the Hamiltonian is conserved along the optimal trajectory, $H=0$ throughout, reducing the minimal action \eqref{eq:mft_path_int} to
\begin{equation}\label{eq:ldf_no_H}
\psi(r(x))=\int_{0}^{1}\mathrm{d}x\int_{-\infty}^0\mathrm{d}t\,\widehat{\rho}\,\partial_t\rho.
\end{equation}

For generic transport coefficients, the coupled nonlinear Euler-Lagrange equations, together with the unequal boundary densities and the non-uniform typical steady-state density profile, make an exact solution evidently formidable. A useful starting point is the equal-reservoir case, $\rho_b=\rho_a$, for which the typical steady-state profile is uniform, $\bar{\rho}(x)=\rho_a$. For a finite bulk drive, however, the average current is finite $\bar{J}_{0}=\frac{\nu}{2}\sigma(\rho_a)$, corresponding to a purely bulk-driven NESS. Remarkably, the Euler-Lagrange equations then admit the exact solution
\begin{equation}\label{eq:eq_res_rhohat}
\widehat{\rho}=\int_{\rho_a}^{\rho_{0}}\mathrm{d}z\frac{2D(z)}{\sigma(z)}\equiv\mu(\rho_{0})-\mu(\rho_a),
\end{equation}
with the density field $\rho_{0}(x,t)$ following
\begin{equation}\label{eq:eq_res_rho}
\partial_t\rho_{0}=-\partial_x\bigg(D(\rho_{0})\partial_x\rho_{0}+\frac{\nu}{2}\sigma(\rho_{0})\bigg).
\end{equation}
This is the time-reversed hydrodynamic relaxation under the driving field, $-\nu$, which reduces to the familiar time reversal of equilibrium relaxation for $\nu=0$~\cite{1953_Onsager_Fluctuations}. A finite bulk drive therefore breaks the equilibrium time-reversal structure, but the resulting ldf from the minimal action in \eqref{eq:ldf_no_H} remains entirely local
\begin{equation}\label{eq:loc_ldf}
\psi_{\rm loc}(r|\rho_a)=\int_0^1\mathrm{d}x\,\big[f(r)-f(\rho_a)-(r-\rho_a)f'(\rho_a)\big].
\end{equation}
Thus, despite carrying a finite average current, the purely bulk-driven NESS exhibits no long-range density correlations.

\prlsection{A local field-transformation} The equal-reservoir solution \eqref{eq:eq_res_rhohat} naturally expresses the response field as the difference of two local chemical potentials. Motivated by this, we extend this structure to unequal reservoir densities by introducing a field $F(x,t)$ through
\begin{equation}\label{eq:rhohat_F_trans}
\widehat{\rho}=\int_{F}^{\rho}\mathrm{d}z\,\frac{2D(z)}{\sigma(z)}\equiv\mu(\rho)-\mu(F).
\end{equation}
The spatial boundary conditions in \eqref{eq:opt_bc} impose $F(0,t)=\rho_a$ and $F(1,t)=\rho_b$, while the initial fixed-point condition gives $F(x,-\infty)=\bar{\rho}(x)$.

In terms of the $(F,\rho)$-fields, we decompose the ldf in \eqref{eq:ldf_no_H} in terms of a local and a non-local contribution as $\psi(r(x))=\psi_{\rm loc}(r(x)|\bar{\rho}(x))+\psi_{\rm nloc}(r(x))$, where the local component takes the form in \eqref{eq:loc_ldf}. The non-local component of the ldf arises from long-range density correlations, which will only generically manifest for unequal reservoir densities. This non-local ldf takes the form
\begin{equation}\label{eq:nloc_ldf}
\psi_{\rm nloc}(r)=-\int_0^1\mathrm{d}x\int_{-\infty}^0\mathrm{d}t\big(f'(F)-f'(\bar{\rho})\big)\partial_t\rho.
\end{equation}
The optimal $(F,\rho)$-fields are solutions of the Euler-Lagrange equations, obtained from \eqref{eq:old_el_eqns},
\begin{subequations}\label{eq:new_el_eqns}
\begin{align}
\partial_tF&=\frac{D(\rho)}{D(F)}\partial_x\big(D(F)\partial_xF\big)-\frac{\nu}{2}\sigma'(\rho)\partial_xF\nonumber\\
&\quad+\frac{D(F)\sigma'(\rho)-D(\rho)\sigma'(F)}{\sigma(F)}(\partial_xF)^2, \label{eq:new_el_i}\\
\partial_t\rho&=-\partial_x\bigg[D(\rho)\partial_x\rho+\sigma(\rho)\bigg(\frac{\nu}{2}-\frac{2D(F)}{\sigma(F)}\partial_xF\bigg)\bigg]. \label{eq:new_el_ii}
\end{align}
\end{subequations}

The origin of non-locality is now explicit: the field $F$ defines a local chemical-potential landscape $\mu(F)=f'(F)$ whose evolution couples the optimal density dynamics across space. If $F(x,t)$ remains locked to $\bar\rho(x)$, the integrand in the RHS of \eqref{eq:nloc_ldf} vanishes identically, so that $\psi_{\rm nloc}(r)=0$ and the LDF remains local despite the boundary drive.

\prlsection{Exactly solvable cases} Equation~\eqref{eq:new_el_i} reveals two distinct routes for which $F(x,t)=\bar{\rho}(x)$ remains invariant throughout the optimal trajectory. In both cases, the stationary $F$-field makes the LDF local, $\psi(r(x))=\psi_{\rm loc}(r(x)|\bar{\rho}(x))$, but the associated optimal fluctuation dynamics are markedly different.

The first arises when the bulk driving field exactly balances the boundary-induced chemical-potential gradient, $\nu=\mu(\rho_b)-\mu(\rho_a)$. The steady-state current then vanishes, $\bar{J}=0$, and $D(\bar{\rho})\bar{\rho}'=\frac{\nu}{2}\sigma(\bar{\rho})$, so that $F=\bar{\rho}$ satisfies \eqref{eq:new_el_i} for arbitrary $D(\rho)$ and $\sigma(\rho)$. The optimal density evolution consequently reduces to
\begin{equation}
\partial_t\rho=-\partial_x\bigg(D(\rho)\partial_x\rho-\frac{\nu}{2}\sigma(\rho)\bigg),
\end{equation}
which is precisely the time reversal of the hydrodynamic relaxation.

A second route arises when the transport coefficients satisfy $\frac{\sigma'(\rho)}{D(\rho)}={\rm constant}$, a class that includes, for example, Zero-Range Processes and Random-Average Processes. The last term in \eqref{eq:new_el_i} then vanishes identically, and $F=\bar{\rho}$ solves the resulting $F$-equation for arbitrary values of $\nu$. Using $D(\bar{\rho})\bar{\rho}'=\frac{\nu}{2}\sigma(\bar{\rho})-\bar{J}$, the optimal density evolves as
\begin{equation}
\partial_t\rho=-\partial_x\bigg(D(\rho)\partial_x\rho-\frac{\nu}{2}\sigma(\rho)+2\bar{J}\frac{\sigma(\rho)}{\sigma(\bar{\rho})}\bigg).
\end{equation}
The additional $\bar{J}$-dependent term makes the optimal fluctuation generically distinct from time-reversed hydrodynamic relaxation whenever the steady-state current is finite.

A distinct exactly solvable class arises for constant diffusivity and quadratic mobility, encompassing familiar models such as the Simple-Exclusion Process and the Kipnis-Marchioro-Presutti model. For this class, the MFT equations possess a remarkable connection to classical integrability~\cite{2022_Bettelheim_Inverse,2022_Mallick_Exact,2024_Bettelheim_Complete,2026_Krajenbrink_Integrability} and can, in principle, be solved using inverse-scattering methods, although obtaining explicit results for the present problem remains challenging. Here, however, the exact solution follows directly from the $F$-field transformation, without recourse to inverse-scattering or other machinery of classical integrability. In contrast to the stationary-$F$ cases discussed above, the $F$-field here evolves nontrivially, giving rise to a genuinely non-local LDF. Nevertheless, the structure of \eqref{eq:new_el_i} allows further simplification. It can be rearranged into a form for which setting the two sides to zero separately yields
\begin{equation}\label{eq:F_quad}
\partial_tF=-\partial_x\bigg(D\partial_xF-\frac{\nu}{2}\sigma(F)\bigg),
\end{equation}
together with
\begin{equation}\label{eq:rho_quad}
\rho=F-\frac{2}{\sigma''}\frac{\sigma(F)}{\partial_xF}\bigg[\frac{\partial_x^2F}{\partial_xF}+\partial_x\ln\bigg(1-\frac{\nu}{2D}\frac{\sigma(F)}{\partial_xF}\bigg)\bigg].
\end{equation}
These relations automatically satisfy the second Euler-Lagrange equation \eqref{eq:new_el_ii}, as well as the boundary conditions. The coupled optimal trajectory therefore reduces to determining $F(x,t)$ from \eqref{eq:F_quad}, with $\rho(x,t)$ then fixed directly by \eqref{eq:rho_quad}. Evaluating \eqref{eq:ldf_no_H} along this trajectory then yields the exact LDF,
\begin{align}
\psi(r)&=\psi_{\rm loc}(r|F)-\frac{4D}{\sigma''}\int_{0}^{1}\mathrm{d}x\,\big(\varphi(F)-\varphi(\bar{\rho})\big),
\end{align}
where
\begin{equation}
\varphi(F)=\ln F'+\bigg(1-\frac{2D}{\nu}\frac{F'}{\sigma(F)}\bigg)\ln\bigg(1-\frac{\nu}{2D}\frac{\sigma(F)}{F'}\bigg),
\end{equation}
and the profile $F(x)$ is determined implicitly by
\begin{equation}
r=F-\frac{2}{\sigma''}\frac{\sigma(F)}{F'}\bigg[\frac{F''}{F'}+\partial_x\ln\bigg(1-\frac{\nu}{2D}\frac{\sigma(F)}{F'}\bigg)\bigg].
\end{equation}
Along with the boundary conditions $F(0)=\rho_a$ and $F(1)=\rho_b$, this relation determines $F(x)$ as a functional of the entire prescribed profile $r(x)$. The non-locality therefore arises from this non-local dependence of $F$ on $r$. Eliminating $F(x)$ renders $\psi(r(x))$ a genuinely non-local functional of the density profile, encoding the long-range correlations of the steady state.

\prlsection{Perturbative results} A broad class of systems, including exclusion processes with extended particles and interacting Brownian particles~\cite{2004_Schonherr_Exclusion,2005_Lin_From,2005_Schonherr_Hard,2013_Krapivsky_Dynamics,2026_Grabsch_Macroscopic,2026_Saha_A,2026_Jangid_An}, lies outside these exactly solvable cases. The $F$-transformation nevertheless provides a systematic perturbative route for generic $D(\rho)$ and $\sigma(\rho)$. We consider a weak boundary drive $\epsilon\equiv\rho_a-\rho_b\ll1$ at fixed bulk drive $\nu$, and expand the optimal fields as $F(x,t)=\rho_a+\sum_{k\ge1}\epsilon^kF_k(x,t)$ and $\rho(x,t)=\rho_{0}(x,t)+\sum_{k\ge1}\epsilon^k\rho_k(x,t)$. 
Since the non-local contribution vanishes for equal reservoirs, $\epsilon=0$, we similarly write
\begin{equation}
\psi_{\rm nloc}(r)=\sum_{k\ge1}\epsilon^k\psi_k(r).
\end{equation}
Expanding the Euler-Lagrange equations \eqref{eq:new_el_eqns} together with their boundary conditions determines the optimal fields recursively in powers of $\epsilon$, which can then be substituted into \eqref{eq:nloc_ldf} to obtain the non-local LDF order by order. We restrict ourselves here to the main results, with details provided in the Supplemental Material~\cite{Supp_Mat}.

At linear order, $\psi_1(r)=-f''(\rho_a)\int_{0}^{1}\mathrm{d}x\int_{-\infty}^{0}\mathrm{d}t\left(F_1-\bar{\rho}_1\right)\partial_t\rho_0$, where $F_1(x,t)$ follows $\partial_tF_1=D(\rho_0)\partial_x^2F_1-\frac{\nu}{2}\sigma'(\rho_0)\partial_xF_1$. For finite $\nu$, the evolution of $\rho_0(x,t)$ generically drives $F_1$ away from its stationary value $F_1(x,-\infty)=\bar{\rho}_1(x)$, producing a non-local contribution already at $\mathrm{O}(\epsilon)$. By contrast, for $\nu=0$, $F_1(x,t)=\bar{\rho}_1(x)$ remains stationary and $\psi_1$ vanishes identically. The leading non-local contribution therefore shifts to $\mathrm{O}(\epsilon^2)$, $\psi_2(r)=-f''(\rho_a)\int_{0}^{1}\mathrm{d}x\int_{-\infty}^{0}\mathrm{d}t\left(F_2-\bar{\rho}_2\right)\partial_t\rho_0$. Thus, a finite bulk driving field shifts the onset of non-locality from quadratic to linear order in the boundary drive.

These leading-order contributions in the driven and undriven cases can be written compactly as
\begin{subequations}
\begin{equation}\label{eq:nloc_ldf_lead_bdry_drive}
\psi_{\rm nloc}(r)\simeq\epsilon\int_{0}^{1}\mathrm{d}x\,\mathcal{C}(x)\int_{0}^{\infty}\mathrm{d}t\int_{\rho_a}^{q(x,t)}\mathrm{d}z\,\Omega(z),
\end{equation}
where
\begin{equation}\label{eq:omega_defn}
\Omega(z)=\frac{D(z)\sigma'(\rho_a)-D(\rho_a)\sigma'(z)}{\sigma(\rho_a)},
\end{equation}
and
\begin{numcases}
{\mathcal{C}(x)=}
\displaystyle\nu^2\frac{\sigma'(\rho_a)}{2D(\rho_a)}\frac{\mathrm{e}^{\nu\frac{\sigma'(\rho_a)}{2D(\rho_a)}x}}{\mathrm{e}^{\nu\frac{\sigma'(\rho_a)}{2D(\rho_a)}}-1},& for $\nu\neq0$,\\
\displaystyle\epsilon f''(\rho_a),& for $\nu=0$.
\end{numcases}
Here, $q(x,t)$ is the equal-reservoir relaxation path evolving from $r(x)$ at $t=0$ to $\rho_a$ at $t=\infty$, under the reversed bulk drive $-\nu$
\begin{equation}\label{eq:eq_reserv_q_relax}
\partial_tq=\partial_x\bigg(D(q)\partial_xq+\frac{\nu}{2}\sigma(q)\bigg).
\end{equation}
\end{subequations}
For $\nu=0$, this reduces to the purely diffusive relaxation.

A few comments are in order. First, for the special case without a bulk drive ($\nu=0$), the leading-order perturbative result in $\bar{J}$ was recently obtained by Bodineau and Derrida in \cite{2025_Bodineau_A}. Second, for systems with a constant $\frac{\sigma'(\rho)}{D(\rho)}$, the $\Omega(z)$-term \eqref{eq:omega_defn} vanishes identically. Third, the leading non-local behavior is entirely described by how a fluctuation relaxes under the reversed bulk drive in the absence of boundary drive.

\prlsection{Long-range correlations} To make explicit how the long-range nature of steady-state correlations is encoded in the non-local LDF, we expand the relaxation path $q(x,t)$ in powers of the fluctuation $\Delta r(x)\equiv r(x)-\rho_a$ and write \eqref{eq:nloc_ldf_lead_bdry_drive} as
\begin{align}
\psi_{\rm nloc}(r)&\simeq\sum_{n=2}^{\infty}\frac{f''(\rho_a)^n}{n!}\bigg[\prod_{i=1}^{n}\int_{0}^{1}\mathrm{d}x^{(i)}\Delta r(x^{(i)})\bigg]\nonumber\\
&\qquad\times K_n\big(\{x^{(i)}\}\big).
\end{align}
Here $-K_n(\{x^{(i)}\})$ gives the leading boundary-drive contribution to the genuinely long-range part of the connected $n$-point density correlation. As an explicit example, for $n=2$ we obtain
\begin{equation}\label{eq:lr_corr_driven}
\mathcal{K}_2(x,y)=\epsilon\frac{\Omega'(\rho_a)}{2f''(\rho_a)^2}\mathcal{C}(x)\int_{0}^{\infty}\mathrm{d}t\,\mathcal{G}(x,t;y,0),
\end{equation}
The Green function $\mathcal{G}$ satisfies the linearized advection-diffusion equation
\begin{equation}\label{eq:green_func_rhoeq}
\partial_t\mathcal{G}(x,t;y,0)=\bigg(D(\rho_a)\partial_x^2+\frac{\nu}{2}\sigma'(\rho_a)\partial_x\bigg)\mathcal{G}(x,t;y,0),
\end{equation}
with vanishing spatial boundary conditions and $\mathcal{G}(x,0;y,0)=\delta(x-y)$. When $\nu=0$, the advective term vanishes and $\mathcal{G}$ is simply the Green's function of the linearized diffusion equation.

Since $\mathcal{G}(x,t;y,0)$ is proportional to the two-time density correlation $C_t^{0}(x,y)$ of the equal-reservoir system~\cite{1990_Garrido_Long,2016_Sadhu_Correlations}, the long-range steady-state correlation can be viewed as the cumulative effect of these spatially extended dynamical correlations over their slow hydrodynamic relaxation, $K_2(x,y)\propto\int_{0}^{\infty}\mathrm{d}t\,C_t^{0}(x,y)$. The same interpretation extends to higher-point correlations, with the explicit three-point ($n=3$) result given in the Supplemental Material~\cite{Supp_Mat}.

The hierarchy of powers of $\Delta r$ in the non-local LDF is controlled by the nonlinearity of the relaxation dynamics \eqref{eq:eq_reserv_q_relax}. For $\nu=0$, constant diffusivity makes this relaxation linear, so that a polynomial mobility of degree $n$ yields a non-local LDF terminating at $O(\Delta r^n)$, whereas density-dependent diffusivity generically generates correlations of arbitrary order. A finite bulk drive removes this truncation even for constant $D$ when $\sigma$ is nonlinear, and thus generically produces an infinite hierarchy of long-range connected multipoint correlations already at linear order in the boundary drive.

\prlsection{Higher dimensions} Our $F$-field transformation within MFT extends directly to a $d$-dimensional slab with reservoirs coupled to two opposite faces~\cite{1983_Spohn_Long}. We distinguish a bulk drive parallel to the boundary-induced density gradient, referred to as longitudinal driving, from one orthogonal to it, referred to as transverse driving. The class with constant $\frac{\sigma'(\rho)}{D(\rho)}$ remains exactly solvable for either orientation, with a local LDF, while the zero-current solution is possible only for longitudinal driving. By contrast, the non-local exact solution for constant-$D$ and quadratic-$\sigma$ does not persist beyond the effectively one-dimensional sector. For generic transport coefficients, the drive orientation controls the onset of non-locality: at linear order, the bulk-advection term drives $F_1(\mathbf{x},t)$ away from its stationary profile $\bar{\rho}(x_1)$ for longitudinal driving, while it vanishes for transverse driving. Consequently, non-locality appears at linear order in the boundary drive in the longitudinal case, but only at quadratic order in the transverse case, as in the undriven case. Details are given in the End Matter.

\prlsection{Conclusion} In this \emph{Letter}, we have shown that the transformation to the $F$-field provides a unified formalism for understanding density-profile fluctuations in boundary-driven diffusive systems through the structure of their optimal dynamics. In particular, the evolution of $F$ distinguishes local from non-local fluctuation costs and reveals when the optimal trajectory departs from time-reversed hydrodynamic relaxation. Beyond the known exactly solvable cases, our perturbative results show that bulk driving qualitatively reshapes steady-state correlations, shifting the leading non-local contribution to linear order in the boundary drive and, generically, producing long-range connected correlations of arbitrarily high order. The same picture persists in higher dimensions, where the effect of the bulk drive depends on its orientation relative to the boundary-induced density gradient.

Several open directions emerge from our results. An important question is whether the $F$-field can be used to characterize Lagrangian phase transitions~\cite{2010_Bertini_Lagrangian,2012_Bunin_Non,2016_Shpielberg_Le} arising from competing optimal trajectories. It would also be interesting to extend the $F$-formalism to periodic geometries~\cite{2008_Appert-Rolland_Universal,2013_Espigares_Dynamical}, where boundary driving is absent. Finally, the strongly biased limit of weakly bulk-driven systems~\cite{2004_Enaud_Large,2009_Bertini_Strong} offers a natural route to understanding how the present diffusive description crosses over toward ballistic dynamics~\cite{2025_Berlioz_Tracer,2023_Doyon_Ballistic}. Extending the $F$-transformation to systems with multiple coupled hydrodynamic fields~\cite{2024_Han_Scaling,2024_Meerson_Relaxation}, as arise in active matter~\cite{2021_Agranov_Exact,2024_Mukherjee_Hydrodynamics}, and to systems with multiple conserved quantities~\cite{2016_Bertini_Transport,2016_Castro-Alvaredo_Emergent}, including integrable~\cite{2023_Doyon_Emergence,2026_Kethepalli_Ballistic} and quantum systems~\cite{2021_Bernard_Can,2023_McCulloch_Full,2024_Wienand_Emergence}, would further test the generality of the formalism.

\prlsection{Acknowledgments} This work originated from discussions with Bernard Derrida, who suggested the problem and drew our attention to the transformation to the $F$-variable. TS gratefully acknowledges Bernard Derrida's 2017 Coll\`ege de France lecture notes, which were immensely helpful in shaping this work. We acknowledge financial support from the Department of Atomic Energy, Government of India, under Project Identification No. RTI-4012. TS also acknowledges support from the International Research Project (IRP) titled ``Classical and quantum dynamics in out of equilibrium systems'' by CNRS, France.

\bibliographystyle{apsrev4-2}
\bibliography{letter}

\renewcommand{\theequation}{E.\arabic{equation}}
\setcounter{equation}{0}

\renewcommand{\thefigure}{E.\arabic{figure}}
\setcounter{figure}{0}

\raggedbottom

\addtolength{\abovedisplayskip}{-1pt}
\addtolength{\belowdisplayskip}{-1pt}

\par\bigskip
{\centering\large\bfseries End Matter\par}
\medskip

\prlsection{Exactly solvable models}
In Table~\ref{tab:exact_cases}, we list representative model systems belonging to the two exactly solvable classes identified in the main text. For these models, the $F$-field transformation determines the density LDF together with the optimal trajectory realizing a prescribed density fluctuation.

\begin{table*}[t]
\caption{\label{tab:exact_cases}
Representative models belonging to the two exactly solvable classes discussed in the main text. The first class satisfies
$\sigma'(\rho)/D(\rho)=\mathrm{const.}$~\cite{2009_Bertini_Towards,2015_Bertini_Macroscopic,2016_Sadhu_Correlations},
while the second has constant $D$ and quadratic $\sigma$~\cite{2013_Carinci_Duality,2021_Ayala_Higher,2022_Floreani_Orthogonal}.}
\begin{ruledtabular}
\begin{tabular}{lcc}
\text{Model} & $D(\rho)$ & $\sigma(\rho)$\\
\colrule

\multicolumn{3}{c}{Case I: A constant ${\displaystyle\frac{\sigma'(\rho)}{D(\rho)}}$}\\
\colrule
Ginzburg--Landau Model~\cite{1988_Guo_Nonlinear,1991_Spohn_Large}
& arbitrary & constant\\
Reflecting Brownian Particles~\cite{2009_Derrida_Current2,2014_Krapivsky_Large}
& $1$ & $2\rho$\\
Zero-Range Process~\cite{1970_Spitzer_Interaction,2005_Harris_Current}
& $g'(\rho)$ & $2g(\rho)$\\
Random Average Process~\cite{1998_Ferrari_Fluctuations,2016_Kundu_Exact}
& ${\displaystyle\frac{\mu_1}{2\rho^2}}$
& ${\displaystyle\frac{\mu_1\mu_2}{(\mu_1-\mu_2)\rho}}$\\

\colrule
\multicolumn{3}{c}{Case II: $D(\rho)={\rm constant}$ and $\sigma(\rho)={\rm quadratic}$}\\
\colrule
Simple Exclusion Process~\cite{2007_Derrida_Non,2015_Bertini_Macroscopic}
& $1$ & $2\rho(1-\rho)$\\
Simple Partial Exclusion Process~\cite{1994_Schutz_Non,2023_Franceschini_Hydrodynamical}
& $N$ & $2\rho(N-\rho)$\\
Kipnis--Marchioro--Presutti Model~\cite{1982_Kipnis_Heat,2005_Bertini_Large}
& $1$ & $2\rho^2$\\
Simple Inclusion Process~\cite{2007_Giardinà_Duality,2022_Franceschini_Symmetric}
& $K$ & $2\rho(K+\rho)$\\
Harmonic Process~\cite{2020_Frassek_Non,2025_Carinci_Large}
& ${\displaystyle\frac{1}{s}}$
& ${\displaystyle\frac{2\rho}{s}\Big(1+\frac{\rho}{s}\Big)}$\\
A Dynamical-Phase-Transition Model~\cite{2018_Baek_Dynamical}
& $1$ & $1+\rho^2$
\end{tabular}
\end{ruledtabular}
\end{table*}

\prlsection{Higher-dimensional setup and results} A key strength of MFT is that the framework readily extends to higher dimensions~\cite{2013_Akkermans_Universal,2025_Bodineau_A}. We illustrate this for a $d$-dimensional slab with coordinates $\mathbf{X}=(X_1,\mathbf{X}_\perp)$, where $X_1\in[0,L]$ and $\mathbf{X}_\perp\in\mathbb{R}^{d-1}$, coupled to reservoirs of densities $\rho_a$ and $\rho_b$ at $X_1=0$ and $X_1=L$, respectively. Further, a weak bulk drive $\nu/L$ is applied along the $X_j$-direction, with unit vector $\hat{\mathbf e}_j$. The cases $j=1$ and $j\neq1$ correspond to longitudinal and transverse driving, respectively. On the hydrodynamic scale $(\mathbf{x},t)=(\mathbf{X}/L,T/L^2)$, the coarse-grained density field $\rho(\mathbf{x},t)$ obeys the $d$-dimensional generalization of the fluctuating hydrodynamics in \eqref{eq:fhd_eqn}, with
\begin{equation}
\rho(0,\mathbf{x}_\perp,t)=\rho_a,\quad\rho(1,\mathbf{x}_\perp,t)=\rho_b.
\end{equation}

Translational invariance along the transverse directions implies that the average steady-state profile depends only on $x_1$, $\bar{\rho}(\mathbf{x})=\bar{\rho}(x_1)$. Fluctuations about this profile obey the large-deviation form $\Pr(r(\mathbf{x}))\asymp\exp[-L^d\psi(r(\mathbf{x}))]$. The path-integral formulation in \eqref{eq:mft_defn} extends directly to $d>1$, with the LDF obtained by minimizing the corresponding $d$-dimensional action over histories connecting the typical profile to the prescribed fluctuation $\rho(\mathbf{x},0)=r(\mathbf{x})$. Following the same argument as in one dimension, the Hamiltonian vanishes along the optimal trajectory, reducing the LDF to the $d$-dimensional generalization of \eqref{eq:ldf_no_H}. The optimal fields satisfy the corresponding generalizations of the Euler--Lagrange equations \eqref{eq:old_el_eqns}, with
\begin{equation}
\widehat{\rho}(0,\mathbf{x}_\perp,t)=\widehat{\rho}(1,\mathbf{x}_\perp,t)=0.
\end{equation}

For equal reservoirs, $\rho_b=\rho_a$, the optimal fluctuation dynamics retains the same structure as in one dimension for generic transport coefficients. The response field is still given by \eqref{eq:eq_res_rhohat}, while the density field $\rho_0(\mathbf{x},t)$ follows
\begin{equation}
\partial_t\rho_0=-\boldsymbol{\nabla}\cdot\bigg(D(\rho_0)\boldsymbol{\nabla}\rho_0+\frac{\nu}{2}\sigma(\rho_0)\hat{\mathbf e}_j\bigg).
\end{equation}

We therefore apply the same transformation \eqref{eq:rhohat_F_trans} in $d$ dimensions. The resulting Euler-Lagrange equations are the direct higher-dimensional generalization of \eqref{eq:new_el_eqns}, with the bulk drive directed along $\hat{\mathbf e}_j$. The corresponding spatial boundary conditions for the $F$ field are
\begin{equation}
F(0,\mathbf{x}_\perp,t)=\rho_a,\quad F(1,\mathbf{x}_\perp,t)=\rho_b,
\end{equation}
while the initial fixed-point condition is
\begin{equation}
F(\mathbf{x},-\infty)=\bar{\rho}(x_1).
\end{equation}

In higher dimensions, the exactly solvable structure becomes more restrictive. The class of models with constant $\frac{\sigma'(\rho)}{D(\rho)}$ remains exactly solvable for arbitrary drive orientation, with $F(\mathbf{x},t)=\bar{\rho}(x_1)$ and a local LDF. The zero-current solution likewise persists for arbitrary transport coefficients, but in the slab geometry it can be realized only for a longitudinal drive, since the bulk field must align with the boundary-induced chemical-potential gradient. By contrast, the non-local exact solution for constant $D$ and quadratic $\sigma$ does not extend to generic $d$-dimensional fluctuations. Away from these special cases, we therefore turn to a perturbative treatment in the boundary drive $\epsilon=\rho_a-\rho_b$.

At leading order in the boundary drive, the field $F_1(\mathbf{x},t)$ obeys
\begin{equation}\label{eq:F1_higher_d}
\partial_tF_1=D(\rho_0)\nabla^2F_1-\frac{\nu}{2}\sigma'(\rho_0)\partial_{x_j}F_1.
\end{equation}
For longitudinal driving, $j=1$, the advective term survives and, as $\rho_0$ evolves, disrupts the initial diffusion--advection balance, causing $F_1(\mathbf{x},t)$ to evolve. Consequently, the non-local LDF appears already at $O(\epsilon)$. For transverse driving, $j\neq1$, the leading field depends only on $x_1$, $F_1(\mathbf{x},t)=-x_1$, so that $\partial_{x_j}F_1=0$ and $F_1$ remains stationary. The linear contribution therefore vanishes and non-locality first appears at $O(\epsilon^2)$, as in the undriven case.

Both cases can be written compactly as
\begin{equation}\label{eq:nloc_higher_d}
\psi_{\rm nloc}(r)\simeq\epsilon\int\mathrm{d}\mathbf{x}\,\mathcal{C}(\mathbf{x})\int_0^\infty\mathrm{d}t\int_{\rho_a}^{q(\mathbf{x},t)}\mathrm{d}z\,\Omega(z),
\end{equation}
where
\begin{numcases}
{\mathcal{C}(\mathbf{x})=}
\displaystyle\nu^2\frac{\sigma'(\rho_a)}{2D(\rho_a)}\frac{\mathrm{e}^{\nu\frac{\sigma'(\rho_a)}{2D(\rho_a)}x_1}}{\mathrm{e}^{\nu\frac{\sigma'(\rho_a)}{2D(\rho_a)}}-1},& $j=1$ and $\nu\ne0$,\\
\displaystyle\epsilon f''(\rho_a),& $j\ne1$ or $\nu=0$.
\end{numcases}
Here, $q(\mathbf{x},t)$ denotes the equal-reservoir relaxation path,
satisfying
\begin{equation}\label{eq:q_higher_d}
\partial_tq=\boldsymbol{\nabla}\cdot\bigg(D(q)\boldsymbol{\nabla}q+\frac{\nu}{2}\sigma(q)\hat{\mathbf e}_j\bigg),
\end{equation}
with
\begin{equation}
q(\mathbf{x},0)=r(\mathbf{x}),\quad q(\mathbf{x},t\to\infty)=\rho_a,
\end{equation}
and
\begin{equation}
q(0,\mathbf{x}_\perp,t)=q(1,\mathbf{x}_\perp,t)=\rho_a.
\end{equation}
Since $\mathcal{C}(\mathbf{x})=O(\epsilon)$ for $j\ne1$ with a finite bulk drive, the transverse contribution in \eqref{eq:nloc_higher_d} starts at $O(\epsilon^2)$, whereas the longitudinal contribution is generically already $O(\epsilon)$.

Expanding the leading non-local LDF further in powers of $\Delta r(\mathbf{x})=r(\mathbf{x})-\rho_a$, the quadratic contribution takes the form
\begin{align}\label{eq:psi2_higher_d}
\psi_{\rm nloc}^{(2)}(r)\simeq\epsilon\frac{\Omega'(\rho_a)}{4}\int\mathrm{d}\mathbf{x}\,\mathrm{d}\mathbf{y}&\int_0^\infty\mathrm{d}t\,\Delta r(\mathbf{x})\,\Delta r(\mathbf{y})\,\mathcal{C}(\mathbf{x})\nonumber\\
&\times\,\mathcal{G}(\mathbf{x},t;\mathbf{y},0),
\end{align}
with the kernel
\begin{equation}
K(\mathbf{x},\mathbf{y})=\mathcal{C}(\mathbf{x})\int_0^\infty\mathrm{d}t\,\mathcal{G}(\mathbf{x},t;\mathbf{y},0).
\end{equation}
Here, $\mathcal{G}$ is the Green's function of the linearized longitudinal advection-diffusion operator,
\begin{equation}\label{eq:gr_fu_w_drift}
\partial_t\mathcal{G}(\mathbf{x},t;\mathbf{y},0)=\bigg(D(\rho_a)\nabla_{\mathbf{x}}^2+\frac{\nu}{2}\sigma'(\rho_a)\partial_{x_1}\bigg)\mathcal{G}(\mathbf{x},t;\mathbf{y},0),
\end{equation}
whereas it satisfies
\begin{equation}\label{eq:gr_fu_w_o_drift}
\partial_t\mathcal{G}(\mathbf{x},t;\mathbf{y},0)=D(\rho_a)\nabla_{\mathbf{x}}^2\mathcal{G}(\mathbf{x},t;\mathbf{y},0),
\end{equation}
in absence of a bulk drive.

The Green's function obeys homogeneous Dirichlet conditions at $x_1=0,1$ and
\begin{equation}
\mathcal{G}(\mathbf{x},0;\mathbf{y},0)=\delta^{(d)}(\mathbf{x}-\mathbf{y}).
\end{equation}

For transverse driving, the linearized relaxation itself contains a uniform advection along $x_j$. However, the two propagators entering the quadratic contribution acquire the same transverse displacement, which drops out upon integration over the intermediate spatial coordinate. Equivalently, this common drift can be removed in a comoving frame. Consequently, the quadratic kernel is expressed in terms of the drift-free propagator $\mathcal{G}$ in \eqref{eq:gr_fu_w_o_drift}.

At short separations $s=|\mathbf{x}-\mathbf{y}|\ll1$, away from the boundaries, the drift is subleading and the kernel inherits the singularity of the diffusive Green's function,
\begin{equation}
K(\mathbf{x},\mathbf{y})\propto
\begin{cases}
\displaystyle \ln\frac{1}{s}, & d=2,\\
\displaystyle \frac{1}{s^{d-2}}, & d>2.
\end{cases}
\end{equation}
At large transverse separations $R=|\mathbf{x}_\perp-\mathbf{y}_\perp|\gg1$, the lowest slab mode dominates and the kernel decays exponentially,
\begin{equation}
K(\mathbf{x},\mathbf{y})\asymp\mathrm{e}^{-\kappa R}.
\end{equation}
For longitudinal driving,
\begin{equation}
\kappa=\sqrt{\pi^2+\frac{\nu^2}{16}\frac{\sigma'(\rho_a)^2}{D(\rho_a)^2}},
\end{equation}
whereas
\begin{equation}
\kappa=\pi
\end{equation}
for transverse driving and in the undriven case.

\end{document}


\title{Supplement to ``\emph{Steady-state density large deviations in driven diffusive systems}''}

\author{Soumyabrata Saha}
\email{soumyabrata.saha@tifr.res.in}
\affiliation{Department of Theoretical Physics, Tata Institute of Fundamental Research, Homi Bhabha Road, Mumbai 400005, India}

\author{Tridib Sadhu}
\email{tridib@theory.tifr.res.in}
\affiliation{Department of Theoretical Physics, Tata Institute of Fundamental Research, Homi Bhabha Road, Mumbai 400005, India}

\date{\today}

\begin{abstract}
In this Supplemental Material, we provide details of the results presented in the Letter. We first derive the typical steady-state density profile perturbatively in the boundary drive and discuss the equal-reservoir solution at finite bulk drive. We then derive the exact LDF for the constant-diffusivity, quadratic-mobility class from the corresponding optimal Euler--Lagrange solution. For generic transport coefficients, we present the perturbative construction of the auxiliary field and the resulting non-local LDF at finite bulk drive, together with the undriven case. Finally, we derive the leading long-range two- and three-point steady-state density correlations from the expansion of the non-local LDF.
\end{abstract}

\maketitle

\tableofcontents

\section{Typical steady-state density profile}

In the NESS, the typical density profile $\bar{\rho}(x)$ is determined by the condition of a spatially uniform average current,
\begin{equation}\label{eq:avg_cur_eqn}
\bar{J}=-D(\bar{\rho})\bar{\rho}'+\frac{\nu}{2}\sigma(\bar{\rho}),
\end{equation}
which implies
\begin{equation}\label{eq:typ_den_pro_eqn}
\frac{\mathrm{d}}{\mathrm{d}x}\bigg(D\big(\bar{\rho}(x)\big)\frac{\mathrm{d}\bar{\rho}(x)}{\mathrm{d}x}
-\frac{\nu}{2}\sigma\big(\bar{\rho}(x)\big)\bigg)=0,
\end{equation}
subject to the Dirichlet boundary conditions $\bar{\rho}(0)=\rho_a$ and $\bar{\rho}(1)=\rho_b$.

Introducing the perturbation parameter, $\epsilon\equiv\rho_a-\rho_b$, we expand the typical profile as
\begin{equation}
\bar{\rho}(x)=\rho_a+\sum_{k\ge1}\epsilon^k\bar{\rho}_k(x),
\end{equation}
where we have used that for equal reservoirs ($\rho_b=\rho_a$), the stationary profile is uniform, $\bar{\rho}(x)=\rho_a$. Solving recursively in $\epsilon$, the leading-order contribution is
\begin{equation}\label{eq:rhobar1_driven}
\bar{\rho}_1(x)=-\frac{\mathrm{e}^{\Lambda x}-1}{\mathrm{e}^{\Lambda}-1},
\end{equation}
where, we have denoted
\begin{equation}
\Lambda=\nu\frac{\sigma'(\rho_a)}{2D(\rho_a)}.
\end{equation}

In the zero-bulk-drive limit, $\nu\to0$, this reduces to $\bar{\rho}_1(x)=-x$, independently of the transport coefficients. For $\nu=0$, the next-to-leading contribution is
\begin{equation}\label{eq:rhobar2_undriven}
\bar{\rho}_2(x)=\frac{D'(\rho_a)}{2D(\rho_a)}\,x(1-x),
\end{equation}
which vanishes for constant-diffusivity models.

\section{The equal-reservoir case}
Using the average current in the steady state \eqref{eq:avg_cur_eqn}, we write
\begin{equation}
\bar{\rho}'(x)=\frac{1}{D\big(\bar{\rho}(x)\big)}\bigg(\frac{\nu}{2}\sigma\big(\bar{\rho}(x)\big)-\bar{J}\bigg)
\end{equation}
Since this autonomous first-order equation admits only monotonic non-constant solutions, the equal boundary conditions $\bar{\rho}(0)=\bar{\rho}(1)=\rho_a$ require the uniform profile $\bar{\rho}(x)=\rho_a$ with the finite stationary current $\bar J_{0}=\frac{\nu}{2}\sigma(\rho_a)$.

Using the fluctuation-dissipation relation, we consider the solution
\begin{equation}\label{eq:rhohat0_soln}
\widehat{\rho}(x,t)=\int_{\rho_a}^{\rho_0(x,t)}\mathrm{d}z\,\frac{2D(z)}{\sigma(z)}=f'(\rho_0(x,t))-f'(\rho_a).
\end{equation}
Substitution into the Euler-Lagrange equation of $\rho$ gives
\begin{equation}
\partial_t\rho_0=-\partial_x\bigg(D(\rho_0)\partial_x\rho_0+\frac{\nu}{2}\sigma(\rho_0)\bigg),
\end{equation}
while the Euler-Lagrange equation for $\widehat{\rho}$ is satisfied identically. Thus, $\rho_0$ evolves from $\rho_a$ at $t\to-\infty$ to the prescribed profile
$r(x)$ at $t=0$.

Using the solution \eqref{eq:rhohat0_soln} in the zero-Hamiltonian form of the action,
\begin{align}
\psi(r)&=\int_0^1\mathrm{d}x\int_{-\infty}^{0}\mathrm{d}t\,\widehat{\rho}\,\partial_t\rho_0\nonumber\\
&=\int_0^1\mathrm{d}x\int_{-\infty}^{0}\mathrm{d}t\big(f'(\rho_0)-f'(\rho_a)\big)\partial_t\rho_0,
\end{align}
the time integral can be performed directly to give the local LDF
\begin{align}
\psi(r(x))&=\int_0^1\mathrm{d}x\,\big[f(r(x))-f(\rho_a)-(r(x)-\rho_a)f'(\rho_a)\big]\\
&\equiv\psi_{\rm loc}(r(x)|\rho_a).
\end{align}

\section{Ldf for constant diffusivity and qudratic mobility}
For constant diffusivity, ($D\equiv D(\rho)={\rm constant}$), and quadratic mobility, ($\sigma''\equiv\sigma''(\rho)={\rm constant}$), the Euler-Lagrange equations admit the solution
\begin{subequations}\label{eq:quad_sigma_EL_sol}
\begin{align}
\partial_tF&=-\partial_x\bigg(D\partial_xF-\frac{\nu}{2}\sigma(F)\bigg),\label{eq:quad_mob_soln1}\\
\rho&=F-\frac{2}{\sigma''}\frac{\sigma(F)}{\partial_xF}\bigg[\frac{\partial_x^2F}{\partial_xF}+\partial_x\ln\bigg(1-\frac{\nu}{2D}\frac{\sigma(F)}{\partial_xF}\bigg)\bigg].\label{eq:quad_mob_soln2}
\end{align}
\end{subequations}

To evaluate the LDF along this optimal trajectory, we write
\begin{equation}\label{eq:quad_mob_LDF_interim}
\psi(r)=\psi_{\rm loc}(r(x)|F(x))+\int_{0}^{1}\mathrm{d}x\int_{-\infty}^{0}\mathrm{d}t\,(\rho-F)f''(F)\partial_tF,
\end{equation}
with the local component
\begin{equation}
\psi_{\rm loc}(r|F)=\int_0^1\mathrm{d}x\big[f(\rho)-f(F)-(\rho-F)f'(F)\big].
\end{equation}
Here, we have used the temporal conditions $F(x,-\infty)=\rho(x,-\infty)=\bar{\rho}(x)$ and $\rho(x,0)=r(x)$, and the terminal auxiliary profile $F(x)\equiv F(x,0)$ is determined by
\begin{equation}
r=F-\frac{2}{\sigma''}\frac{\sigma(F)}{F'}\bigg[\frac{F''}{F'}+\partial_x\ln\bigg(1-\frac{\nu}{2D}\frac{\sigma(F)}{F'}\bigg)\bigg],
\end{equation}
subject to $F(0)=\rho_a$ and $F(1)=\rho_b$.

Using the fluctuation-dissipation relation, we rewrite \eqref{eq:quad_mob_soln2} as
\begin{equation}
(\rho-F)f''(F)=-\frac{4D}{\sigma''}\frac{1}{\partial_xF}\partial_x\ln\bigg(\partial_xF-\frac{\nu}{2D}\sigma(F)\bigg).
\end{equation}
Then, the non-local term in the LDF \eqref{eq:quad_mob_LDF_interim} is expressed as
\begin{equation}
\int_{0}^{1}\mathrm{d}x\int_{-\infty}^{0}\mathrm{d}t\,(\rho-F)f''(F)\partial_tF=-\frac{4D}{\sigma''}\int_{0}^{1}\mathrm{d}x\int_{-\infty}^{0}\mathrm{d}t\,\bigg\{\partial_t\varphi(F)+\partial_x\bigg[\frac{2D}{\nu}\frac{\partial_tF}{\sigma(F)}\ln\bigg(1-\frac{\nu}{2D}\frac{\sigma(F)}{\partial_xF}\bigg)\bigg]\bigg\},
\end{equation}
where
\begin{equation}
\phi(F)=\ln\partial_xF+\left(1-\frac{2D}{\nu}\frac{\partial_xF}{\sigma(F)}\right)\ln\left(1-\frac{\nu}{2D}\frac{\sigma(F)}{\partial_xF}\right).
\end{equation}
Since $F(x,t)$ is fixed at the spatial boundaries, the total spatial derivative does not contribute. Performing the time integration, with $(F,\rho)\to(\bar{\rho},\bar{\rho})$ as $t\to-\infty$ and $(F,\rho)\to(F,r)$ at $t=0$, we obtain
\begin{equation}
\psi(r)=\psi_{\rm loc}(r(x)|F(x))-\frac{4D}{\sigma''}\int_0^1\mathrm{d}x\,\big[\varphi(F(x))-\varphi(\bar{\rho}(x))\big].
\end{equation}

\section{Solution for $F_1$ at a finite bulk drive}
We introduce the homogeneous evolution operator
\begin{equation}\label{eq:L_nu_defn}
\mathbb{L}_{\nu}=D(\rho_{0})\partial_x^2-\frac{\nu}{2}\sigma'(\rho_{0})\partial_x.
\end{equation}
Then, at leading order in the boundary drive, the auxiliary field obeys
\begin{equation}\label{eq:F1_evolution}
\partial_tF_1=\mathbb{L}_{\nu}F_1,
\end{equation}
with the spatial and temporal boundary conditions
\begin{equation}
F_1(0,t)=0,\quad F_1(1,t)=-1,\quad F_1(x,-\infty)=\bar{\rho}_1(x),
\end{equation}
where $\bar{\rho}_1(x)$ is given in \eqref{eq:rhobar1_driven}.

The solution of $F_1(x,t)$ can be written as
\begin{equation}\label{eq:F1_solution}
F_1(x,t)=-\frac{\mathrm{e}^{\Lambda x}-1}{\mathrm{e}^{\Lambda}-1}-\frac{\nu}{f''(\rho_a)}\int_{0}^{1}\mathrm{d}y\int_{-\infty}^{t}\mathrm{d}s\,\mathbb{G}_{\nu}(x,t;y,s)\frac{\Lambda\mathrm{e}^{\Lambda y}}{\mathrm{e}^{\Lambda}-1}\,\Omega\big(\rho_{0}(y,s)\big),
\end{equation}
where we have defined
\begin{equation}\label{eq:omega_defn}
\Omega(z)=\frac{D(z)\sigma'(\rho_a)-D(\rho_a)\sigma'(z)}{\sigma(\rho_a)}.
\end{equation}
Here, $\mathbb{G}_{\nu}$ is the Green's function associated with the homogeneous evolution operator $\mathbb{L}_{\nu}$ and satisfies
\begin{equation}\label{eq:green_nu_defn}
\partial_t\mathbb{G}_{\nu}(x,t;y,s)=\bigg(D\big(\rho_{0}(x,t)\big)\partial_x^2-\frac{\nu}{2}\sigma'\big(\rho_{0}(x,t)\big)\partial_x\bigg)\mathbb{G}_{\nu}(x,t;y,s)
\end{equation}
subject to
\begin{equation}
\mathbb{G}_{\nu}(0,t;y,s)=\mathbb{G}_{\nu}(1,t;y,s)=0,\quad\mathbb{G}_{\nu}(x,s;y,s)=\delta(x-y).
\end{equation}

\subsection{Solution for $F_2$ in the absence of bulk drive}
In the absence of a bulk drive ($\nu=0$), the linear-order auxiliary field in \eqref{eq:F1_solution} reduces to its stationary value $F_1(x,t)=\bar{\rho}_1(x)=-x$. The first nontrivial evolution of the auxiliary field therefore appears at $O(\epsilon^2)$. At this order,
\begin{equation}\label{eq:F2_evolution}
\partial_tF_2=\mathbb{L}_0F_2+\frac{D'(\rho_a)}{D(\rho_a)}D(\rho_{0})-\Omega(\rho_{0}),
\end{equation}
subject to
\begin{equation}
F_2(0,t)=F_2(1,t)=0,\quad F_2(x,-\infty)=\bar{\rho}_2(x),
\end{equation}
with $\bar{\rho}_2(x)$ given in \eqref{eq:rhobar2_undriven}. The homogeneous operator is the $\nu=0$ limit of \eqref{eq:L_nu_defn},
\begin{equation}
\mathbb{L}_0=D(\rho_{0})\partial_x^2.
\end{equation}

The solution of $F_2(x,t)$ is then written as
\begin{equation}\label{eq:F2_solution}
F_2(x,t)=\frac{D'(\rho_a)}{2D(\rho_a)}x(1-x)-\int_{0}^{1}\mathrm{d}y\int_{-\infty}^{t}\mathrm{d}s\,\mathbb{G}_{0}(x,t;y,s)\,\Omega\big(\rho_{0}(y,s)\big)
\end{equation}
with the $\Omega$ as defined in \eqref{eq:omega_defn}. Here, $\mathbb{G}_{0}$ is the Green's function associated with $\mathbb{L}_0$, satisfying
\begin{equation}
\partial_t\mathbb{G}_{0}(x,t;y,s)=D\big(\rho_{0}(x,t)\big)\partial_x^2\mathbb{G}_{0}(x,t;y,s),
\end{equation}
with
\begin{equation}
\mathbb{G}_{0}(0,t;y,s)=\mathbb{G}_{0}(1,t;y,s)=0,\quad\mathbb{G}_{0}(x,s;y,s)=\delta(x-y).
\end{equation}

\section{Solution of $\psi_1$ at a finite bulk drive}
Starting from \eqref{eq:F1_evolution}, we write for $\delta F_1(x,t)\equiv F_1(x,t)-\bar{\rho}_1(x)$,
\begin{equation}\label{eq:F1_evo}
\partial_t\delta F_1=\mathbb{L}_\nu\delta F_1-\frac{\nu}{f''(\rho_a)}\frac{\Lambda\mathrm{e}^{\Lambda x}}{\mathrm{e}^\Lambda-1}\Omega(\rho_{0}),
\end{equation}
Similarly, differentiating the equal-reservoir optimal density evolution with respect to time gives
\begin{equation}\label{eq:rhoeq_evo}
\partial_t(\partial_t\rho_{0})=-\mathbb{L}_\nu^\dagger(\partial_t\rho_{0}),
\end{equation}
where $\mathbb{L}_\nu^\dagger$ is the adjoint operator given by
\begin{equation}
\mathbb{L}_{\nu}^\dagger\Phi=\partial_x^2\big(D(\rho_{\rm eq})\Phi\big)+\frac{\nu}{2}\partial_x\big(\sigma'(\rho_{\rm eq})\Phi\big).
\end{equation}

The linear-order coefficient of the non-local LDF is
\begin{equation}
\psi_1(r)=-f''(\rho_a)\int_{-\infty}^0\mathrm{d}t\,I_1(t),
\end{equation}
where we have defined
\begin{equation}
I_1(t)=\int_0^1\mathrm{d}x\,\delta F_1(x,t)\,\partial_t\rho_{\rm eq}(x,t).
\end{equation}
Using the evolution equations of $\delta F_1$ \eqref{eq:F1_evo} and $\partial_t\rho_{\rm eq}$ \eqref{eq:rhoeq_evo}, and integrating by parts, the homogeneous terms cancel. Since both $\delta F_1$ and $\partial_t\rho_{\rm eq}$ vanish at the boundaries, we get
\begin{equation}
\frac{\mathrm{d}I_1(t)}{\mathrm{d}t}=-\frac{\nu}{f''(\rho_a)}\,\frac{\Lambda}{\mathrm{e}^\Lambda-1}\int_0^1\mathrm{d}x\,\mathrm{e}^{\Lambda x}\,\Omega(\rho_{\rm eq})\,\partial_t\rho_{\rm eq}.
\end{equation}
Using the identity
\begin{equation}\label{eq:omega_derivative_identity}
\Omega(\rho_{\rm eq})\,\partial_t\rho_{\rm eq}=\partial_t\int_{\rho_a}^{\rho_{\rm eq}}dz\,\Omega(z),
\end{equation}
and the fixed-point conditions $\delta F_1(x,-\infty)=0$ and $\rho_{\rm eq}(x,-\infty)=\rho_a$, we obtain
\begin{equation}
I_1(t)=-\frac{\nu}{f''(\rho_a)}\,\frac{\Lambda}{\mathrm{e}^\Lambda-1}\int_0^1\mathrm{d}x\,\mathrm{e}^{\Lambda x}\int_{\rho_a}^{\rho_{\rm eq}(x,t)}\mathrm{d}z\,\Omega(z).
\end{equation}
Finally, introducing the forward relaxation
$q(x,t)=\rho_{\rm eq}(x,-t)$ for $0<t<\infty$, we arrive at the announced result
\begin{align}
\psi_{\rm nloc}(r)&=\epsilon\psi_1(r)+\mathrm{O}(\epsilon^2)\nonumber\\
&\simeq\epsilon\frac{\nu\Lambda}{\mathrm{e}^\Lambda-1}\int_0^1\mathrm{d}x\,\mathrm{e}^{\Lambda x}\int_0^\infty\mathrm{d}t\int_{\rho_a}^{q(x,t)}\mathrm{d}z\,\Omega(z).
\end{align}

\subsection{Solution of $\psi_2$ in the absence of bulk drive}
For $\nu=0$, $F_1(x,t)=\bar\rho_1(x)=-x$ remains stationary throughout the optimal trajectory. Hence $\delta F_1=0$ and $\psi_1(r)=0$. The leading non-local contribution therefore appears at quadratic order in the boundary drive, through a nontrivially evolving $F_2(x,t)$. Using \eqref{eq:F2_evolution}, we write for $\delta F_2(x,t)\equiv F_2(x,t)-\bar{\rho}_2(x)$
\begin{equation}
\partial_t\delta F_2=\mathbb{L}_0\delta F_2-\Omega(\rho_{\rm eq}).
\end{equation}

In this undriven case, the quadratic-order coeffcient of the non-local LDF is
\begin{equation}
\psi_2(r)=-f''(\rho_a)\int_{-\infty}^0\mathrm{d}t\,I_2(t),
\end{equation}
with
\begin{equation}
I_2(t)=\int_0^1\mathrm{d}x\,\delta F_2(x,t)\,\partial_t\rho_{\rm eq}(x,t).
\end{equation}

Proceeding as before, we take a time-derivative where the homogeneous terms cancel, and then use \eqref{eq:omega_derivative_identity} to write
\begin{equation}
I_2(t)=-\int_0^1\mathrm{d}x\int_{\rho_a}^{\rho_{\rm eq}(x,t)}\mathrm{d}z\,\Omega(z).
\end{equation}
Since the $\mathrm{O}(\epsilon)$ non-local LDF vanishes, we write to the leading order in $\epsilon$,
\begin{align}
\psi_{\rm nloc}(r)&=\epsilon^2\psi_2(r)+\mathrm{O}(\epsilon^3)\nonumber\\
&\simeq\epsilon^2f''(\rho_a)\int_0^1\mathrm{d}x\int_0^\infty\mathrm{d}t\int_{\rho_a}^{q(x,t)}\mathrm{d}z\,\Omega(z).
\end{align}

\section{The optimal density trajectories}
For a finite bulk drive at the linear order in the boundary drive, the density follows
\begin{equation}
\partial_t\rho_1=-\mathbb{L}_{\nu}^\dagger\rho_1+f''(\rho_a)\partial_x\big(\sigma(\rho_0)\partial_xF_1\big),
\end{equation}
with the boundary conditions
\begin{equation}
\rho_1(0,t)=0,\quad\rho_1(1,t)=1,\quad\rho_1(x,-\infty)=\bar{\rho}_1(x).
\end{equation}
When the bulk drive is zero, this reduces to
\begin{equation}
\partial_t\rho_1=-\mathbb{L}_0^\dagger\rho_1-f''(\rho_a)\partial_x\big(\sigma(\rho_0)\big),
\end{equation}
where we used $F_1(x,t)=-x$ and replaced $\mathbb{L}_{\nu}^\dagger$ with $\mathbb{L}_0^\dagger$.

\section{Long-range steady-state density correlations}
We expand the leading non-local LDF in powers of the final profile fluctuation around the typical profile of the equal-reservoir system
\begin{equation}
\Delta r(x)\equiv r(x)-\rho_a.
\end{equation}
The equal-reservoir relaxation path entering the non-local LDF is correspondingly expanded as
\begin{equation}\label{eq:q_expansion}
q(x,t)-\rho_a=q_1(x,t)+q_2(x,t)+\mathrm{O}(\Delta r^3),
\end{equation}
where $q_n\equiv\mathrm{O}(\Delta r^n)$. Using $\Omega(\rho_a)=0$, we obtain
\begin{equation}\label{eq:nloc_LDF_expansion}
\psi_{\rm nloc}(r)\simeq\epsilon\int_{0}^{1}\mathrm{d}x\,\mathcal{C}(x)\int_{0}^{\infty}\mathrm{d}t\,\bigg(\frac{\Omega'(\rho_a)}{2}q_1^2+\Omega'(\rho_a)q_1q_2+\frac{\Omega''(\rho_a)}{6}q_1^3\bigg)+\mathrm{O}(\Delta r^4)
\end{equation}
The first term in the parenthesis gives the quadratic contribution to the non-local LDF,
whereas the remaining two determine the cubic contribution.

To determine $q_1$ and $q_2$, we expand the relaxation equation
\begin{equation}
\partial_tq=\partial_x\bigg(D(q)\partial_xq+\frac{\nu}{2}\sigma(q)\bigg).
\end{equation}
For the sake of brevity, we define
\begin{equation}
\mathcal{L}_\nu\equiv D(\rho_a)\partial_x^2+\frac{\nu}{2}\sigma'(\rho_a)\partial_x.
\end{equation}
the first two orders satisfy
\begin{align}
(\partial_t-\mathcal{L}_\nu)\,q_1&=0,\label{eq:q1_evolution}\\
(\partial_t-\mathcal{L}_\nu)\,q_2&=\bigg(\frac{D'(\rho_a)}{2}\partial_x^2+\frac{\nu\sigma''(\rho_a)}{4}\partial_x\bigg)\,q_1^2,\label{eq:q2_evolution}
\end{align}
with the initial conditions
\begin{align}
&q_1(x,0)=\Delta r(x),\nonumber\\
&q_2(x,0)=0,
\end{align}
and homogeneous spatial boundary conditions. Introducing the Green function $\mathcal{G}_\nu$ satisfying
\begin{equation}
\partial_t\mathcal{G}_\nu(x,t;y,0)=\mathcal{L}_\nu\mathcal{G}_\nu(x,t;y,0),
\end{equation}
with
\begin{equation}
\mathcal{G}_\nu(0,t;y,0)=\mathcal{G}_\nu(1,t;y,0)=0,\quad\mathcal{G}_\nu(x,0;y,0)=\delta(x-y),
\end{equation}
we obtain
\begin{equation}\label{eq:q1_solution}
q_1(x,t)=\int_{0}^{1}\mathrm{d}y\,\mathcal{G}_\nu(x,t;y,0)\Delta r(y),
\end{equation}
and
\begin{equation}\label{eq:q2_solution}
q_2(x,t)=\int_{0}^{1}\mathrm{d}y\int_{0}^{t}\mathrm{d}s\,\mathcal{G}_\nu(x,t;y,s)\,\bigg(\frac{D'(\rho_a)}{2}\partial_y^2+\frac{\nu\sigma''(\rho_a)}{4}\partial_y\bigg)\,q_1(y,s)^2.
\end{equation}

At quadratic order in $\Delta r$, the non-local LDF \eqref{eq:nloc_LDF_expansion} therefore becomes
\begin{equation}\label{eq:psi2_intermediate}
\psi_{\rm nloc}^{(2)}(r)=\epsilon\frac{\Omega'(\rho_a)}{2}\int_0^1\mathrm{d}x\,\mathrm{d}y\,\Delta r(x)\,\Delta r(y)\int_0^1\mathrm{d}u\,\mathcal{C}(u)\int_0^\infty\mathrm{d}t\,\mathcal{G}_\nu(u,t;x,0)\,\mathcal{G}_\nu(u,t;y,0).
\end{equation}
The Green's function obeys the identity
\begin{equation}\label{eq:weighted_semigroup}
\int_0^1\mathrm{d}u\,\mathrm{e}^{\Lambda u}\mathcal{G}_\nu(u,t;x,0)\mathcal{G}_\nu(u,t;y,0)=\mathrm{e}^{\Lambda x}\mathcal{G}_\nu(x,2t;y,0)=\mathrm{e}^{\Lambda y}\mathcal{G}_\nu(y,2t;x,0).
\end{equation}
Using this, and rescaling $2t\to t$, \eqref{eq:psi2_intermediate}
reduces to
\begin{equation}
\psi_{\rm nloc}^{(2)}(r)=\epsilon\frac{\Omega'(\rho_a)}{4}\int_0^1\mathrm{d}x\,\mathrm{d}y\,\Delta r(x)\,\Delta r(y)\,\mathcal{C}(x)\int_0^\infty\mathrm{d}t\,\mathcal{G}_\nu(x,t;y,0).
\end{equation}
This yields the $2$-point kernel
\begin{equation}
\mathcal{K}_2(x,y)=\epsilon\frac{\Omega'(\rho_a)}{2f''(\rho_a)^2}\mathcal{C}(x)\int_0^\infty\mathrm{d}t\,\mathcal{G}_\nu(x,t;y,0),
\end{equation}
as announced in the \emph{Letter}.

We next turn to the cubic-in-$\Delta r$ contribution. The $q_1^3$ term follows simply from \eqref{eq:q1_solution}. For the $q_1q_2$ term, consider first the contribution in which $x$ originates from $q_1$, and $y$ and $z$ from $q_2$,
\begin{align}
\int_0^1\mathrm{d}u\int_0^\infty\mathrm{d}t&\,\mathcal{C}(u)\,q_1(u,t)\,q_2(u,t)=\int_0^1\mathrm{d}u\int_0^1\mathrm{d}v\int_0^\infty\mathrm{d}t\int_0^t\mathrm{d}s\,\mathcal{C}(u)\,\mathcal{G}_\nu(u,t;x,0)\nonumber\\
&\times\mathcal{G}_\nu(u,t;v,s)\,\bigg(\frac{D'(\rho_a)}{2}\partial_v^2+\frac{\nu\sigma''(\rho_a)}{4}\partial_v\bigg)\,\big[\mathcal{G}_\nu(v,s;y,0)\,\mathcal{G}_\nu(v,s;z,0)\big]
\end{align}
Using time-translation invariance $\mathcal{G}_\nu(x,t;y,s)=\mathcal{G}_\nu(x,t-s;y,0)$ along with the identity \eqref{eq:weighted_semigroup} the above simplifies to
\begin{align}
\int_0^1\mathrm{d}u\int_0^\infty\mathrm{d}t&\,\mathcal{C}(u)\,q_1(u,t)\,q_2(u,t)=\frac{1}{2}\int_0^1\mathrm{d}v\int_0^\infty\mathrm{d}s\int_s^\infty\mathrm{d}\tau\,\mathcal{C}(x)\,\mathcal{G}_\nu(x,\tau;v,0)\nonumber\\
&\times\bigg(\frac{D'(\rho_a)}{2}\partial_v^2+\frac{\nu\sigma''(\rho_a)}{4}\partial_v\bigg)\,\big[\mathcal{G}_\nu(v,s;y,0)\,\mathcal{G}_\nu(v,s;z,0)\big]
\end{align}
where we have changed the time variable $\tau=2t-s$. Finally performing an integration by parts and combining with the $q_1^3$ term we arrive at the $\mathrm{O}(\Delta r^3)$ term
\begin{align}
\psi_{\rm nloc}^{(3)}(r)=\frac{f''(\rho_a)^3}{6}\int_0^1\mathrm{d}x\,\mathrm{d}y\,\mathrm{d}z\,\Delta r(x)\,\Delta r(y)\,\Delta r(z)\,\mathcal{K}_3(x,y,z),
\end{align}
with the $3$-point kernel
\begin{align}
&\mathcal{K}_3(x,y,z)=\frac{\epsilon}{f''(\rho_a)^3}\int_{0}^{1}\mathrm{d}u\int_{0}^{\infty}\mathrm{d}t\,\Bigg[\Omega''(\rho_a)\,\mathcal{C}(u)\,\mathcal{G}_\nu(u,t;x,0)\mathcal{G}_\nu(u,t;y,0)\mathcal{G}_\nu(u,t;z,0)\nonumber\\
&+\Omega'(\rho_a)\sum_{{\rm cyc}(x,y,z)}\mathcal{C}(x)\,\mathcal{G}_\nu(u,t;y,0)\,\mathcal{G}_\nu(u,t;z,0)\,\bigg(\frac{D'(\rho_a)}{2}\partial_u^2-\frac{\nu\sigma''(\rho_a)}{4}\partial_u\bigg)\,\mathcal{H}(x,t;u)\Bigg].
\end{align}
Here, we denoted $\mathcal{H}(x,t;u)=\int_{t}^{\infty}\mathrm{d}s\,\mathcal{G}_\nu(x,s;u,0)$ and the cyclic combination $\sum_{{\rm cyc}(x,y,z)}F(x,y,z)=F(x,y,z)+F(y,z,x)+F(z,x,y)$ ensures the full permutation symmetry of the kernel.

For vanishing bulk drive, $\nu=0$, the leading non-local contribution is of $O(\epsilon^2)$, with $\mathcal{G}_\nu\to\mathcal{G}_0$. For constant diffusivity, the relaxation dynamics is linear, so that $q_{n\geq2}=0$ and the multipoint contributions arise solely from powers of $q_1$. In particular, the two- and three-point kernels derived above reduce, after a trivial rescaling of time to set $D=1$, to the corresponding constant-diffusivity results of Bodineau and Derrida.